\documentstyle[multicol,prl,aps,epsf]{revtex}
\begin{document}

\draft

\title{Anti-Kondo resonance in transport through a quantum wire 
       with a side-coupled quantum dot}
\author{Kicheon Kang$^1$\cite{kang}, Sam Young Cho$^2$, Ju-Jin Kim$^1$
and Sung-Chul Shin$^3$}
\address{   $^1$Department of Physics, Chonbuk National University,
            Chonju 561-756, Chonbuk, Korea}

\address{   $^2$Institute of Physics and Applied Physics, Yonsei University,
            Seoul 120-749, Korea}

\address{ $^3$Department of Physics, 
          Korea Advanced Institute of Science and Technology,
          Taejon 305-701, Korea }  

\date{\today}

\maketitle

\begin{abstract}
An interacting quantum dot side-coupled to a perfect quantum wire
is studied. Transport through the quantum wire is investigated
by using an exact sum rule and the slave-boson mean field treatment. 
It is shown that the Kondo effect provides a {\em suppression} of
the transmission due to the destructive interference of the ballistic
channel and the Kondo channel. At finite temperatures, anti-resonance
behavior is found as a function of the quantum dot level position,
which is interpreted as a crossover from the high temperature Kondo
phase to the low temperature charge fluctuation phase.
\end{abstract}
%
\begin{multicols}{2}
The Kondo effect has been a subject of intensive study for more than
three decades~\cite{hewson93}. Initially the problem
was studied to explain anomalous behavior of magnetic impurities
embedded in a metallic host. Recent progress
in the nano-fabrication technique of electronic devices enabled 
investigation of the Kondo effect by means of quantum dots (QD) 
in a very controllable way~\cite{gordon98,cronen98,schmid98,simmel99}.  
The Kondo effect in the single electron transistor (SET)
structures is characterized by an enhanced conductance due to the Kondo 
resonance~\cite{glazman88,ng88,hershfeld,meir,yeyati93,konig96,kang98,craco99}. 
The Kondo resonance at the Fermi level provides a new channel for electric
current flowing through the QD. On the other hand, it is interesting
to note that the Kondo scattering is a phase-coherent phenomenon 
which is characterized by a resonance phase shift of $\pi/2$ with
unitary tunneling cross section at low temperature and 
bias~\cite{glazman88,ng88}. The phase coherence of the Kondo scattering
cannot be proved
by the transport measurement in SET structures because the conductance
through the QD measures only the probability of the transmission amplitude.
In order to prove the phase-coherence one needs a device setup which
includes quantum interference process. A typical example 
is a QD embedded in an Aharonov-Bohm 
interferometer~\cite{bruder96,davidovich97,izumida97,gerland00,ji00} 
or in an Aharonov-Bohm ring~\cite{ferrari99,kang00,eckle00}, where the
phase-coherence is studied through the dependence on the Aharonov-Bohm
flux. 

In this paper, we consider another type of 
simple quantum interference device with
Kondo correlation: A perfect quantum wire with a side-coupled quantum
dot (Fig.1(a)). This structure is reminiscent of T-shaped quantum wave
guides known as electron stub tuners~\cite{stub} in which the
quasi-bound states in the stub play a major role in transport properties.
On the other hand, we consider relatively weak tunneling limit between the wire
and the dot. Therefore, both electron-electron interaction
and quantum interference are important.
In contrast to the SET geometry, transmission through the system
consists of the interference between the ballistic channel
and the resonant channel from the quantum dot, and the transport
contains some informations on the phase-coherence through
the quantum dot. We consider
a simple one-dimensional quantum wire with a side-coupled Anderson
impurity at site ``0" (Fig.1(b)). 
The Hamiltonian of the system can be written as
\begin{mathletters}
\begin{equation}
 H = H_0 + H_D + H_T ,
\end{equation}
where $H_0$, $H_D$, and $H_T$ represent the noninteracting wire,
the interacting QD, and tunneling,
respectively, which are given by
\begin{eqnarray}
 H_0 &=& \sum_{k\sigma} \varepsilon_k c^\dagger_{k\sigma} c_{k\sigma} \;, \\
 H_D &=& \sum_\sigma \varepsilon_d d_\sigma^\dagger d_\sigma
	         + U \hat{n}_\uparrow \hat{n}_\downarrow \; , \\
 H_T &=& -t' \sum_\sigma \left( d^\dagger_\sigma
	       c_{0\sigma} + c_{0\sigma}^\dagger d_\sigma \right) . 
\end{eqnarray}
\end{mathletters}
Here $\varepsilon_d$ and $U$ represent the single-particle energy
and the on-site Coulomb repulsion in the QD, respectively. $t'$
denotes the tunneling matrix element between the QD level and
the site ``$0$" of the quantum wire. Note that the electron operator
at site $0$ is given by
\begin{equation}
 c_{0\sigma} = \frac{1}{\sqrt{N}} \sum_k c_{k\sigma} ,
\end{equation}
where $N$ is the number of lattice sites that goes to infinity.

At low temperature and bias voltage, electron transport is dominated by
the coherent transmission, and the linear-response conductance
is given by the Landauer-type formula
\begin{equation}
 G = \frac{2e^2}{h} \int \left(-\frac{\partial f}{\partial\varepsilon}
     \right) T(\varepsilon) d\varepsilon \; ,
\end{equation}
where $f$ and $T(\varepsilon)$ denote the Fermi distribution function 
and the transmission probability of an electron with energy
$\varepsilon$, respectively. $T(\varepsilon)$ is related to the
Green's function at site 0 with spin $\sigma$, $G_{0\sigma}$:
\begin{equation}
 T(\varepsilon) = \Gamma^2 \left| G_{0\sigma}(\varepsilon) \right|^2
    \; , \label{eq:trans}
\end{equation}
where $\Gamma$ corresponds to the coupling strength of the site 0
to the other part of the wire (which is proportional to the kinetic
energy of the electrons in the wire).
$G_{0\sigma}$ can be rewritten in terms
of the ``exact" Green function of the dot, $G_{d\sigma}$,
\begin{equation}
 G_{0\sigma}(\varepsilon) = \frac{1}{\varepsilon+i\Gamma}
    \left( 1 + \frac{t'^2}{\varepsilon+i\Gamma} G_{d\sigma}(\varepsilon)
    \right) \; . \label{eq:G0}
\end{equation}
At zero temperature only electrons at the Fermi level, $\varepsilon=0$,
are important, and inelastic processes are fully suppressed. 
In this case the conductance is simplified as
\begin{equation}
 G = \frac{2e^2}{h} T(0) \; , \label{eq:conductance0}
\end{equation}
and the dot Green's function $G_{d\sigma}$ satisfies the Friedel-Langreth
sum rule~\cite{langreth66}:
\begin{equation}
 G_{d\sigma}(0) = \frac{1}{ -\Gamma'\cot{\varphi} + i\Gamma' } \; ,
   \label{eq:sumrule}
\end{equation}
where $\Gamma'= t'^2/\Gamma$ is the resonance width of the quantum dot level,
and 
\begin{equation}
 \varphi = \frac{\pi}{2} n_d \label{eq:phase} 
\end{equation}
with $n_d$ being the average occupation number of the QD.

By combining Eqs.(\ref{eq:trans}), (\ref{eq:G0}), (\ref{eq:conductance0}),
and (\ref{eq:sumrule}), one get the following very compact 
form of the zero-temperature conductance:
\begin{equation}
 G = \frac{2e^2}{h} \cos^2{\varphi} \;.  \label{eq:G-phi}
\end{equation}
The expression (\ref{eq:G-phi}) is in good contrast with the case where the
quantum dot is connected to two separate electrodes. 
In such a case, the zero-temperature
conductance is given by $G=\frac{2e^2}{h} \sin^2{\varphi}$ for a 
symmetrically coupled junctions~\cite{glazman88,ng88}, and
has maximum value in the Kondo limit, $n_d\simeq1$.
However, in our model, the conductance is fully {\em suppressed} in the
Kondo limit. This is purely quantum effect associated with the
Kondo correlation and can be understood in terms of the destructive
interference between the ballistic channel and the Kondo resonance
represented in the first and the second term of Eq.(\ref{eq:G0}), 
respectively.
On the other limit, $n_d=0$ or $n_d=2$, both spin and charge fluctuations
are suppressed. Therefore, the coupling between the quantum
dot and the wire becomes ineffective and the conductance 
approaches to the conductance
quantum due to the ballistic transmission through the wire.

To study the problem in more detail, we adopt the slave-boson mean
field theory (SBMFT) which is known to be a good approximation
for describing the Fermi-liquid properties with strongly renormalized
parameters~\cite{hewson93}. 
In the framework of the SBMFT for infinite $U$, the dot Green's function
is given by 
\begin{equation}
 G_{d\sigma}(\varepsilon) = \frac{1-n_d}{\varepsilon-\tilde{\varepsilon}_d
   +i(1-n_d)\Gamma'} \;, \label{eq:Gd}
\end{equation}
where the renormalized energy level $\tilde{\varepsilon}_d$ and the
occupation number $n_d$ are to be obtained by solving the self-consistent 
equations~\cite{hewson93}:
\begin{mathletters}
\begin{eqnarray}
 \tilde{\varepsilon}_d-\varepsilon_d &=& -\frac{2}{\pi}
   \frac{\partial}{\partial n_d} \int_{-D}^D f(\varepsilon) 
   \tan^{-1}{ \left[ \frac{(1-n_d)\Gamma'}{\tilde{\varepsilon}_d-\varepsilon}
              \right] }
     d\varepsilon  \; ,\\
 n_d &=& -\frac{2}{\pi}\int f(\varepsilon) 
   \Im{ \frac{1}{\varepsilon-\tilde{\varepsilon}_d+ i(1-n_d)\Gamma'} } 
     d\varepsilon  \; .
\end{eqnarray}
 \label{eq:coupled-eq}
\end{mathletters}
A constant density of states (DOS) for the wire at $-D<\varepsilon<D$ is
assumed in the above equation for simplicity.
At zero temperature the integrations in Eq.(\ref{eq:coupled-eq}) 
can be performed analytically and we get
\begin{mathletters}
\begin{eqnarray}
 \tilde{\varepsilon}_d-\varepsilon_d &=& -\frac{2\Gamma'}{\pi} 
   \log{ \frac{ \sqrt{\tilde{\varepsilon}_d^2+(1-n_d)^2\Gamma'^2} }{ D } }
      \\
 n_d &=& \frac{2}{\pi} 
         \tan^{-1}{ \frac{ (1-n_d)\Gamma' }{ \tilde{\varepsilon}_d } } \;. 
	 \label{eq:nd}
\end{eqnarray}
\end{mathletters}

A very important point is that
the SBMFT satisfies automatically the unitarity of the scattering matrix,
that is,
\begin{equation}
 R(\varepsilon) + T(\varepsilon) = 1 \; ,
\end{equation}
where $R(\varepsilon)$ and $T(\varepsilon)$ denote the reflection and
the transmission probability of an electron with energy $\varepsilon$,
respectively. It should be noted that some other methods based on
the large-$N_s$ approach ($N_s$ being the spin degeneracy of the dot), 
such as the non-crossing approximation (NCA) or diagrammatic $1/N_s$
expansion, do not satisfy this relation.
In addition, Eqs. (\ref{eq:G0}), (\ref{eq:conductance0}), (\ref{eq:Gd}),
and (\ref{eq:nd}) lead to the expression (\ref{eq:G-phi}). That is,
the zero-temperature conductance obtained by the SBMFT gives the exact
expression of Eq.(\ref{eq:G-phi}). 

Fig.2 displays the conductance as a function of the quantum dot 
energy level. In the experiment $\varepsilon_d$ can be controlled by
attaching a voltage gate to the QD. 
At zero temperature, the conductance shows a crossover
from resonant reflection in the Kondo limit ($\varepsilon_d \ll 
-2\Gamma'$) to perfect transmission where
the coupling to the wire is ineffective ($\varepsilon_d >
2\Gamma'$). The perfect reflection in the Kondo limit originates
from the completely destructive interference between the ballistic and
the Kondo channel. At finite temperatures,
the conductance shows a dip structure with an asymmetric line 
shape~\cite{comment}.
The dip occurs, as a peak does for the Kondo effect
in the SET structure, due to a continuous transition
from the high-temperature Kondo phase to the low-temperature charge
fluctuation phase. 
As the temperature increases the dip position moves
toward $\varepsilon_d=0$ with reduced depth. 

Fig. 3 explains the crossover behavior for a fixed temperature 
$T=0.1\Gamma'$, in relation to the Kondo temperature, $T_K$.
At very low energy level ($\varepsilon_d \ll -2\Gamma'$)
the Kondo temperature is much smaller than the given temperature $T$.
In this case the Kondo effect is negligible and the 
Kondo-resonant reflection is suppressed due to the ineffective coupling
between the QD and the wire. 
By increasing $\varepsilon_d$, the
Kondo temperature is increased exponentially and the Kondo effect becomes
important. Then the current reaches
its minimum at a point that satisfies the condition $T_K>T$
and $\varepsilon_d < -2\Gamma'$ ($n_d\simeq1$). 
Further increasing $\varepsilon_d$ results in
an increase of the current because of charge fluctuation. Kondo physics
is no longer valid in this region and the transmission through the
QD begins to resemble that of a pure quantum wire as $n_d$ decreases
further, again due to the ineffective coupling.

In conclusion, we have analyzed the quantum transport through a
ballistic quantum wire with a side-coupled interacting quantum 
dot. We have shown that the Kondo resonance associated with a quantum
interference leads to a perfect reflection of the electrons. We have
also discussed the anti-resonance behavior which originates from the
crossover of the high temperature Kondo to the low temperature
charge fluctuation region.



%
%
\begin{figure}[h]
\epsfxsize=3.3in
\epsffile{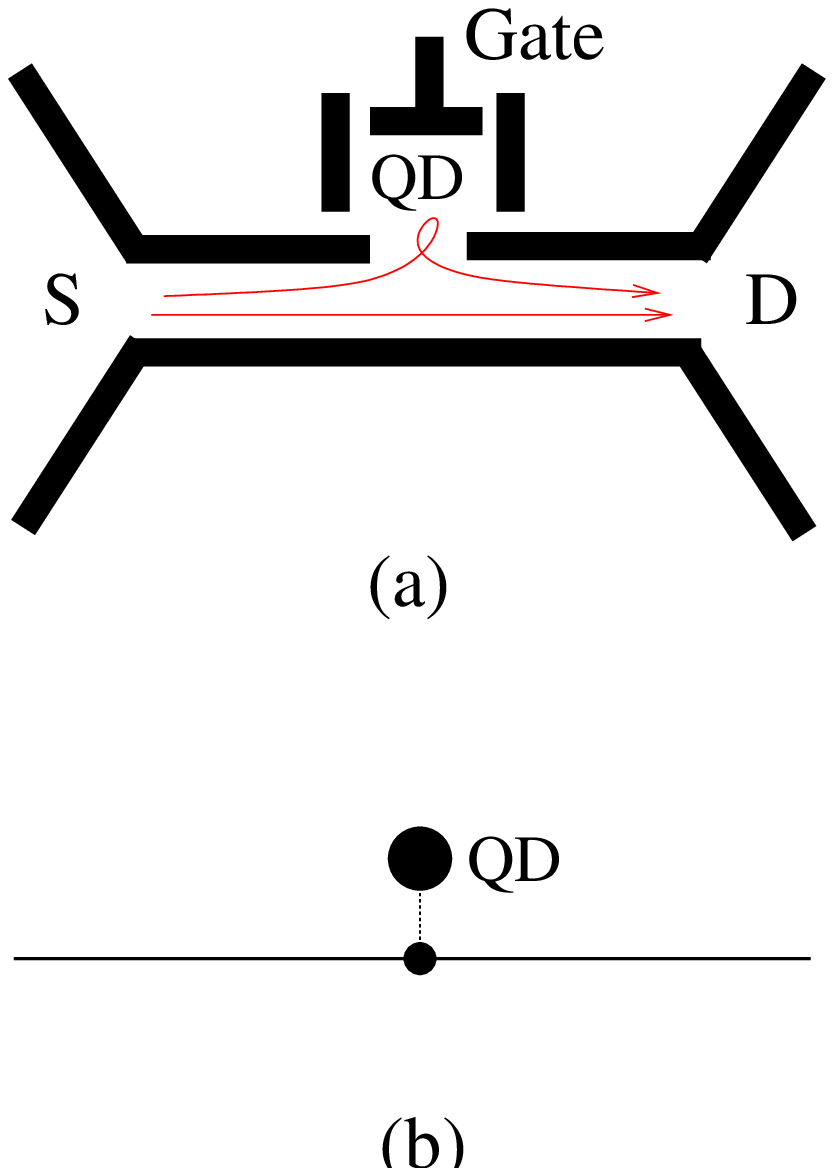}
 \caption{ (a) Schematic diagram of the quantum wire with a side-coupled
  quantum dot.
  (b) Model of the device for the quantum wire and a side-attached quantum
  dot. The quantum wire is considered to be impurity-free one-dimensional 
  metal and the quantum dot is modeled as an Anderson impurity.
	   }
\end{figure}
\begin{figure}[h]
\epsfxsize=3.3in
\epsffile{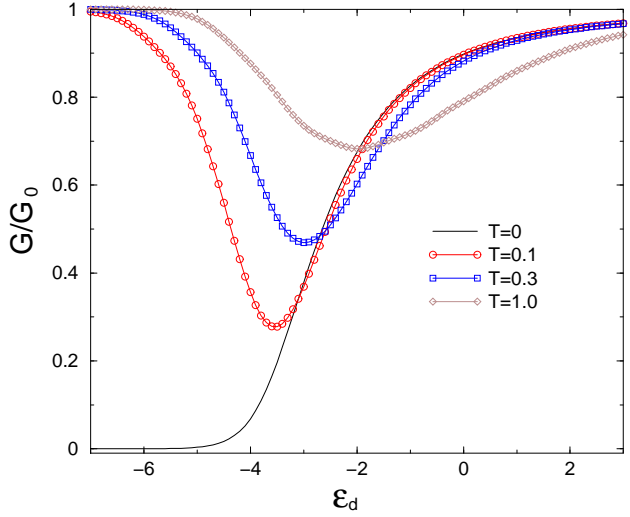}
 \caption{ Conductance (in unit of $G_0=2e^2/h$)
 as a function of the level position of the
 quantum dot ($\varepsilon_d$) for four different temperatures.
 All the energy scale is renormalized in unit of $\Gamma'$ in this figure
 and in Fig. 3.
          }
\end{figure}
\begin{figure}[h]
\epsfxsize=3.3in
\epsffile{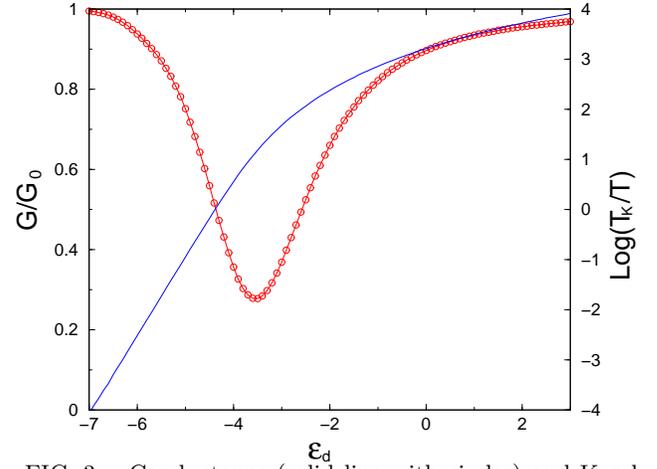}
 \caption{ Conductance (solid line with circles) and Kondo temperature 
 in log scale (solid line), as a function
 of the level position of the quantum dot for $T=0.1\Gamma'$. 
          }
\end{figure}
\end{multicols}
\end{document}